\documentclass[12pt]{article}
\input epsf

\usepackage{a4}
\usepackage{amsfonts,amsmath,amssymb}
\usepackage[pdftex]{graphicx}
\usepackage{bbm}
\usepackage{dsfont}

\newcommand{\nn}{\nonumber}
\newcommand{\Ord}{{\cal O}}
\newcommand{\cA}{{\cal A}}
\newcommand{\cF}{{\cal F}}

\makeatletter

\@addtoreset{equation}{section}
\makeatother

\def\href#1#2{#2}

\begin{document}

\begin{titlepage}
\begin{flushleft}
{Version \today }
\end{flushleft}

\begin{center}

\hfill 
\vskip 1.0in

{\large \bf $U(1)_{B-L}$ extra-natural inflation with Standard Model on a brane}\\[14mm]

{Kazuyuki Furuuchi}$^\dagger$ and
{Jackson M.~S.~Wu}$^\ddagger$
\vskip8mm
{}$^\dagger${\sl Manipal Center for Natural Sciences, Manipal University}\\
{\sl Madhav Nagar, Manipal, Karnataka 576104, India}\\[3mm]
{}$^\ddagger${\sl Physics Division, National Center for Theoretical Sciences}\\
{\sl No.~101, Section 2, Kuang Fu Road, Hsinchu 30013, Taiwan, R.O.C.}\\

\vskip8mm 
\end{center}
\begin{abstract}
The interrelation between inflationary cosmology and new physics beyond the Standard Model (SM) is studied in a $U(1)_{B-L}$ extension of the SM embedded in 
a (4+1)-dimensional spacetime. In the scenario we study, the inflaton arises from the Wilson loop of the $U(1)_{B-L}$ gauge group winding an extra-dimensional cycle. Particular attention is paid to the coupling between the inflaton and SM particles that are confined on a brane localised in the extra dimension.
We find that the inflaton decay channels are rather restricted in this scenario and the resulting reheating temperature is relatively low.
\end{abstract}

\end{titlepage}

\section{Introduction}

The precision of the Cosmic Microwave Background (CMB) anisotropy observations has started to rule out some of the inflation models~\cite{Ade:2013uln}.
However, CMB data alone still accommodates a large class of them. In order to narrow down further likely candidates, it is useful to study possible
relevance of the inflaton to physics in other eras.
In particular, at the time of reheating, the inflaton decays to Standard Model (SM) particles so that the standard hot big bang can proceeed, the nature of the interaction between the inflaton and the SM is thus crucial.

Large field inflation models had attracted attention because of the possibile detection of tensor modes in CMB polarization in the near future~\cite{Lyth:1996im}. It is theoretically challenging to construct natural large field inflation models, since effective field theory approach usually breaks down in these models. Extra-natural inflation~\cite{ArkaniHamed:2003wu,Kaplan:2003aj}, which is based on a gauge theory in higher-dimensional spacetime,
is one way to circumvent this difficulty by using non-local operator (Wilson loop) in the extra dimension.

It is an interesting question what should be the gauge group for extra-natural inflation. As we review in the next section, it turns out that to explain the CMB data the gauge coupling for extra-natural inflation must be very small~\cite{ArkaniHamed:2003wu,Kaplan:2003aj}. 
This makes it difficult to identify the SM gauge groups as that for extra-natural inflation, as their couplings at the electro-weak scale are orders of magnitudes larger than that required for extra-natural inflation.
Therefore we shall look for other gauge groups in models beyond the SM (BSM).

Gauged $U(1)_{B-L}$ extension of the SM~\cite{Mohapatra:1980qe,Marshak:1979fm,Wetterich:1981bx,Masiero:1982fi} is ubiquitous in scenarios of BSM physics. 
A nice feature of it is that the existence of right-handed neutrinos is made natural by the necessity of gauge anomaly cancellation. 
It also makes $R$-parity exact in supersymmetric versions of the SM, and it appears as an intermediate stage in the symmetry breaking pattern of grand-unified models down to the SM, as well as in higher-dimensional embeddings of the SM in string theory constructions. Apart from the formal theoretical considerations, phenomenologically, having a new gauge boson and scalars neutral under the SM gauge group can give rise to novel effects observable in future collider experiments.

In this letter, we study extra-natural inflation with $U(1)_{B-L}$ as the gauge group. In the scenario we study, the bulk spacetime is (4+1)-dimensional with the extra dimension compactified on a circle, SM is confined on a (3+1)-dimensional brane localised in the extra dimension, and the inflation arises from the Wilson loop of the $U(1)_{B-L}$ gauge field living in the full (4+1)-dimensional bulk. 
In the following, 
we explore the interrelation between inflationary cosmology and particle physics in this setting.\footnote{For other approaches to connect inflation and new physics beyond SM via extra-natural inflaton, see~\cite{Inami:2009bs,Inami:2010ke}.} 

The rest of the letter is organised as follows.
The relevant ingredients of extra-natural inflation is reviewed in Sec.~\ref{sec:inflation}.
The details of our $U(1)_{B-L}$ extension of the SM is discussed in Sec.~\ref{sec:BLSM}.
The decay of the inflaton to SM particles is studied in Sec.~\ref{sec:idecay}.
We end with a summary and discussions in Sec.~\ref{sec:summary}.

\section{$U(1)_{B-L}$ extra-natural inflation}\label{sec:inflation}

Extra-natural inflation~\cite{ArkaniHamed:2003wu,Kaplan:2003aj} is a version of natural inflation \cite{Freese:1990rb} whose typical potential takes the form
\begin{eqnarray}
V(\phi) =
\frac{V_0}{2} \left[ 1 - \cos \left( \frac{\phi}{f} \right) \right]\, ,
\label{iL}
\end{eqnarray}
where $\phi$ is the inflaton which, in extra-natural inflation, 
is the zero-mode of the fifth component of some bulk gauge field.
In the scenario we study here, it is that of the $U(1)_{B-L}$ gauge group.
From (\ref{iL}) the slow-roll parameters are given by
\begin{eqnarray}
\label{srpara}
\epsilon_V (\phi)
&\equiv& \frac{M_P^2}{2} \left( \frac{V'}{V} \right)^2 
= \frac{M_P^2}{2f^2}\left( \frac{ \sin \left( \frac{\phi}{f} \right)}{1 - \cos \left( \frac{\phi}{f} \right) } \right)^2  \,,
\label{epsilon} \\
\eta_V (\phi)
&\equiv& M_P^2 \frac{V''}{V} 
= \frac{M_P^2}{f^2} \frac{\cos \left( \frac{\phi}{f} \right)}{1 - \cos \left( \frac{\phi}{f} \right)}
\, .
\label{eta}
\end{eqnarray}
Here $'$ denotes derivative with respect to $\phi$. The slow-roll conditions amount to
\begin{eqnarray}
\epsilon_V,  |\eta_V| \ll 1 \, .
\label{slowrp}
\end{eqnarray}
In extra-natural inflation, $f$ and $V_0$ are estimated as
\cite{ArkaniHamed:2003wu}
\begin{eqnarray}
f  
= \frac{1}{g_{4} (2\pi L_5)}\, ,
\label{f}
\end{eqnarray}
and
\begin{eqnarray}
V_0 
=
\frac{c_0}{\pi^2}
\frac{1}{(2\pi L_5)^4}\, .
\label{V0}
\end{eqnarray}
Here, $g_4$ is the (effective) four-dimensional gauge coupling, and $L_5$ is the radius of the compactified fifth dimension. The constant $c_0$ is determined by the matter content in the bulk, with the relevant ones being fields charged under the gauge symmetry of interest and whose masses are below or of the order of
$1/L_5$~\cite{Hatanaka:1998yp}; each of these field makes an $\Ord(1)$ contribution to $c_0$.\footnote{More precisely, we assume that contributions from charge one fields dominate, which gives rise to the periodicity 
$\phi \sim \phi + 2\pi f$.}

In order for quantum gravity corrections to be small, we need
\begin{eqnarray}
(L_5 M_5)^3 \gg 1\, ,
\label{QG}
\end{eqnarray}
where $M_5$ is the five-dimensional (reduced) Planck scale, which is related to the four-dimensional reduced Planck scale 
$M_P \simeq 2.4 \times 10^{18}$~GeV by
\begin{eqnarray}
M_P^2 = M_5^3 (2 \pi L_5) \,.
\label{M5}
\end{eqnarray}
Thus from~(\ref{f})
\begin{eqnarray}
M_5 = (g_4 f M_P^2)^{1/3}.
\label{M5fg}
\end{eqnarray}
Since $f$ is directly related to the CMB observations, and $g_4$ is a basic parameter in the $U(1)_{B-L}$ extension of the SM, we shall take $f$ and $g_4$ as the independent parameters, and regard $L_5$ and $M_5$ as functions of them. 
It is convenient to introduce a dimensionless parameter
\begin{eqnarray}
\ell_5 
\equiv L_5 M_5 
= \frac{1}{2\pi} \left( \frac{M_P}{g_4 f} \right)^{2/3} \,,
\label{ell5}
\end{eqnarray}
which measures the strength of quantum gravity corrections; 
(\ref{QG}) then amounts to $\ell_5^3 \gg 1$. Although $\ell_5$ is not an independent parameter, it is sometimes convenient to use $\ell_5$ instead of $g_4$. In terms of $\ell_5$ and $f$, $g_4$ is expressed as
\begin{eqnarray}
g_4 = \frac{M_P}{f} (2\pi \ell_5)^{-3/2} \,.
\label{gbound}
\end{eqnarray}

The number of e-folds as a function of $\phi$ is given by
\begin{eqnarray}
N(\phi)
&=&
\int_{t}^{t_e}dt\,
H(\phi(t))
=
\int_{\phi}^{\phi_e} d\phi\,
\frac{H}{\dot{\phi}}
\nn\\
&\simeq&
-\int_{\phi}^{\phi_e} d\phi\,
\frac{1}{M_P^2}\frac{V}{V'}
=
\int^{\phi}_{\phi_e} d\phi\,
\frac{f}{M_P^2}
\frac{1-\cos \frac{\phi}{f}}{\sin \frac{\phi}{f}}
\nn\\
&=&
-
\left(
\frac{f}{M_P}
\right)^2
\log\left[\frac{1}{2}\left( 1 + \cos \frac{\phi}{f} \right)\right]_{\phi_e}^\phi \,.
\label{Nphi}
\end{eqnarray}
Here, $\phi_e$ is the value of the inflaton field at the end of inflation defined by $\epsilon_V (\phi_e) = 1$, where the slow-roll condition (\ref{slowrp}) breaks fown.\footnote{Note that $\epsilon_V \geq |\eta_V|$ for $f > M_P$, which is the case in the following.}
This gives
\begin{eqnarray}
\frac{\phi_e}{f}
=
\cos^{-1} 
\left(
\frac{1-\frac{M_P^2}{2f^2}}{1+\frac{M_P^2}{2f^2}}
\right) \,,
\label{phie}
\end{eqnarray}
and plugging (\ref{phie}) into (\ref{Nphi}) we obtain
\begin{eqnarray}
\frac{\phi}{f} =
\cos^{-1}
\left(
\frac{2 e^{-\frac{M_P^2}{f^2}N}}{1 + \frac{M_P^2}{2f^2}} -1
\right) \,.
\label{phi}
\end{eqnarray}
In slow-roll inflation, the tensor-to-scalar ratio, $r$, and the spectral index, $n_s$, are given by
\begin{eqnarray}
r \simeq 16 \epsilon_V \, , \qquad
n_s \simeq 1 - 6 \epsilon_V + 2 \eta_V \, .
\label{tsns}
\end{eqnarray}
The scalar-to-tensor ratio and the spectral index estimated from various combinations of the Planck data and other observations give at $95\%$ CL:
$r \lesssim 0.12$ and $0.94 \lesssim n_s \lesssim 0.98$ at the pivot scale $k_\ast = 0.002$ Mpc$^{-1}$~\cite{Ade:2013uln}. 
Below, except for $r$ and $n_s$ whose value we take always at the pivot scale, we shall use the subscript $\ast$ to indicate that the value is taken at the pivot scale.

\begin{figure}[htbp]
\centering
\includegraphics[width=5in]{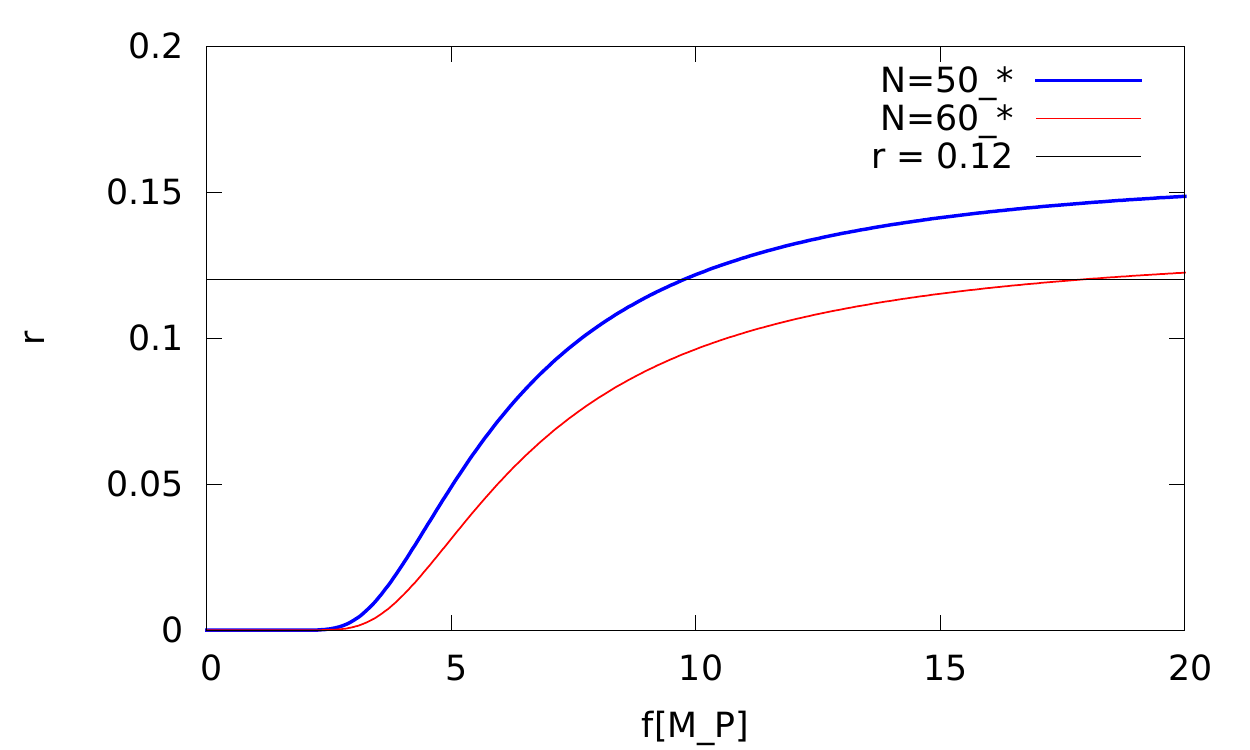} 
\caption{\label{fig:fr} The scalar-to-tensor ratio, $r$, as a function of $f$ for different values of $N_\ast$.}
\end{figure}

\begin{figure}[htbp]
\centering
\includegraphics[width=5in]{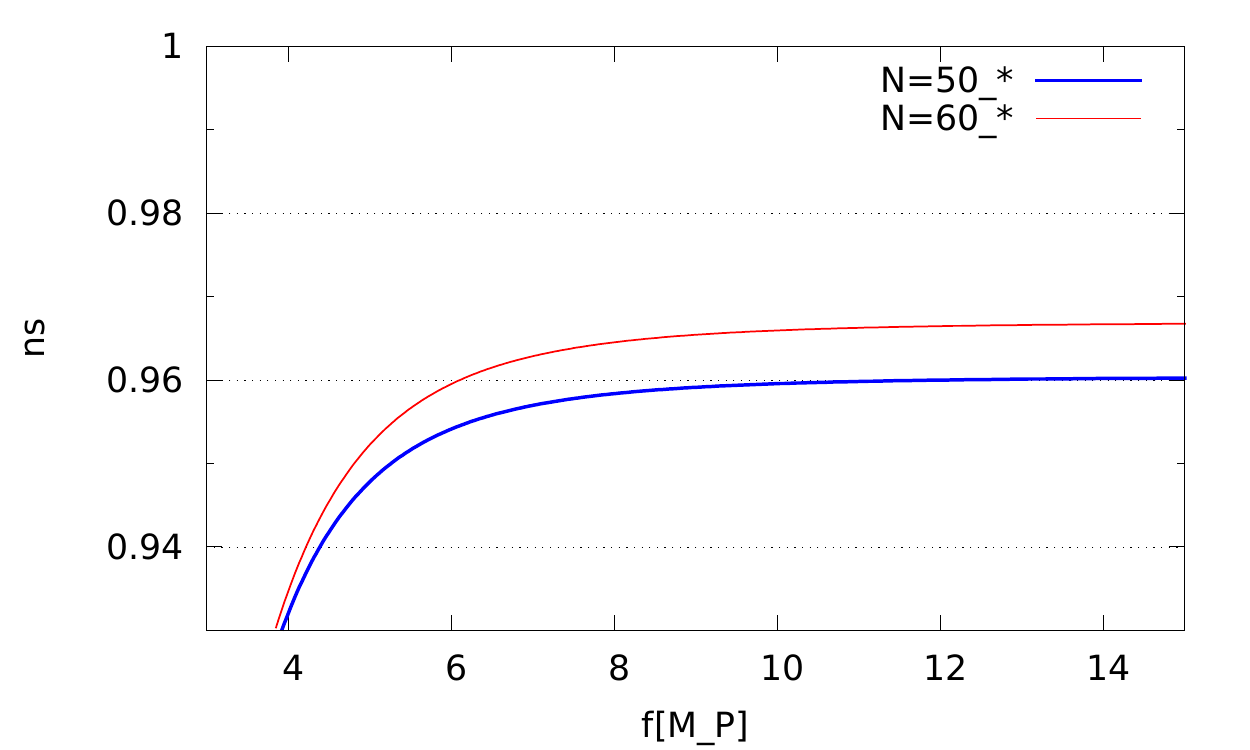} 
\caption{\label{fig:fns} The spectral index, $n_s$, as a function of $f$ for different values of $N_\ast$.}
\end{figure}

As can be seen from (\ref{epsilon}), (\ref{eta}) and (\ref{phi}), $r$ and $n_s$ only depend on $f$ and $N_\ast$ in extra-natural inflation, and so constraints on $r$ and $n_s$ constrain $f$ for a given $N_\ast$. We plot the dependence of $r$ and $n_s$ on $f$ at fixed values of $N_\ast$ in Figs~\ref{fig:fr} and~\ref{fig:fns}, respectively. We see that for $N_\ast=50$, we have $f \lesssim 10M_P$ from $r \lesssim 0.12$ and $f \gtrsim 5M_P$ from $n_s \gtrsim 0.94$.  We will see later when considering the inflaton decay that $N_\ast \simeq 50$ is natural for the scenario we study here.

\begin{figure}[htbp]
\centering
\includegraphics[width=5in]{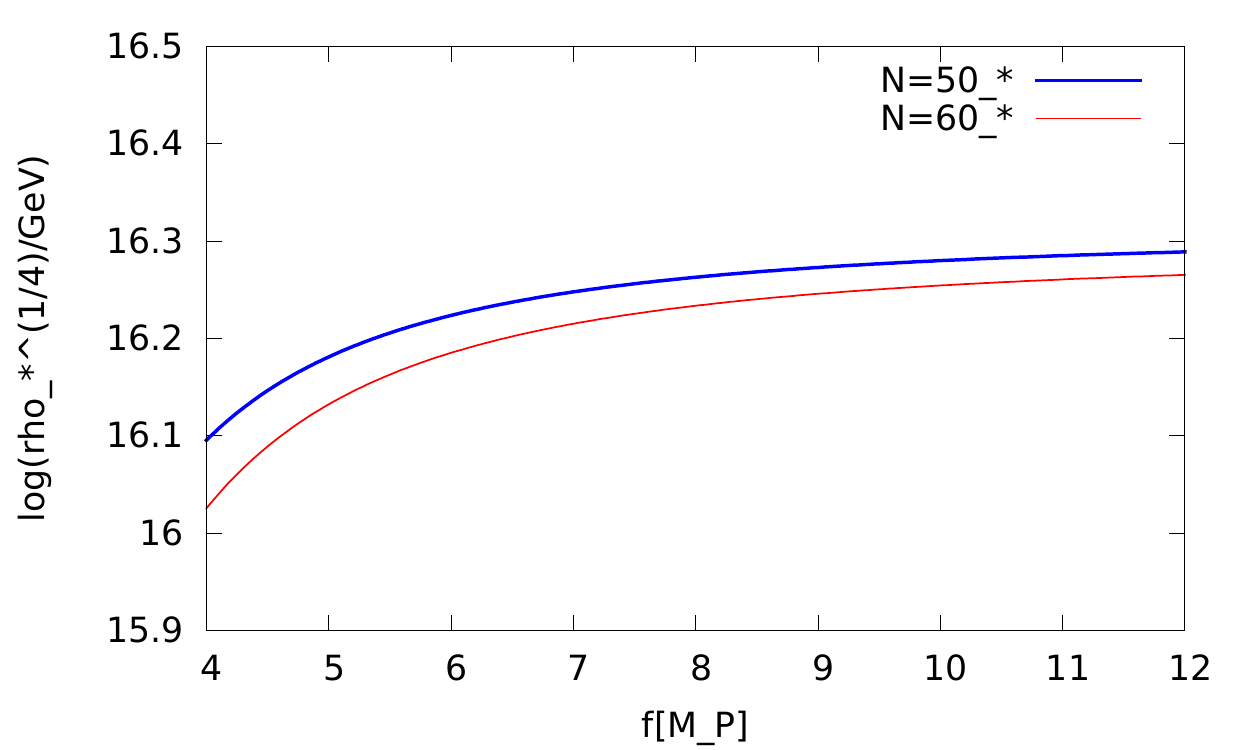} 
\caption{\label{fig:frho} The energy density, $\rho_\ast$, as a function of $f$ for different values of $N_\ast$.}
\end{figure}

\begin{figure}[htbp]
\centering
\includegraphics[width=5in]{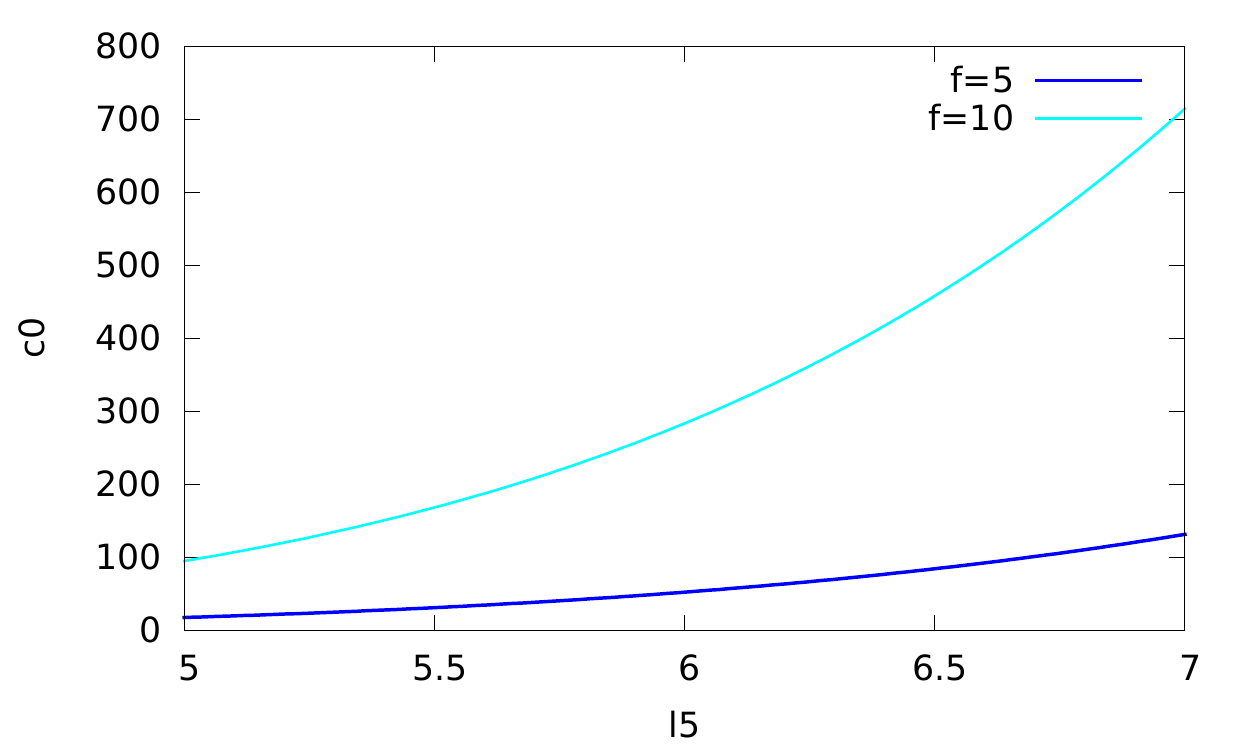} 
\caption{\label{fig:ellc0} The behaviour of $c_0$ as a function of $\ell_5$ for $N_\ast = 50$ at different values of $f$.}
\end{figure}

\begin{figure}[htbp]
\centering
\includegraphics[width=5in]{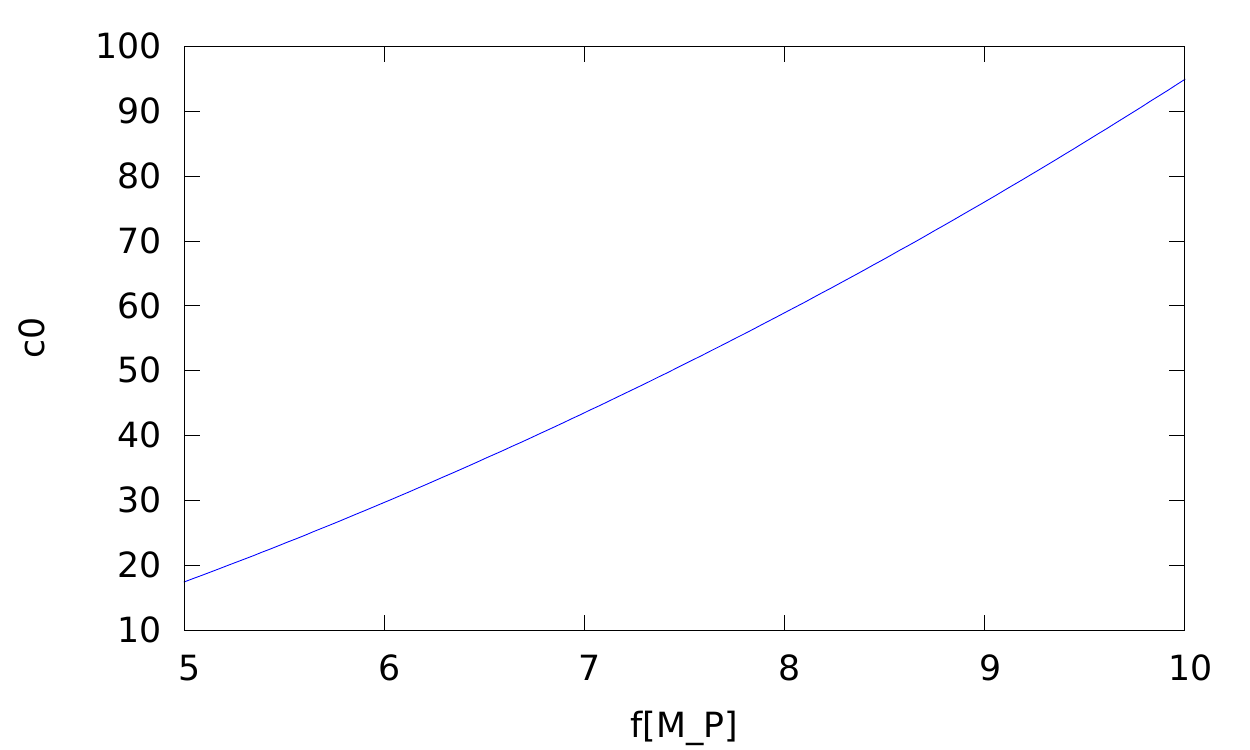} 
\caption{\label{fig:fc0}  The behaviour of $c_0$ as a function of $f$ for $N_\ast=50$ and $\ell_5=5$.}
\end{figure}

The power spectrum of the slow-roll inflation is given by
\begin{eqnarray}
P_\zeta \simeq \frac{H^2}{8\pi^2 M_P^2 \epsilon_V} \, .
\label{PS}
\end{eqnarray}
This should be compared with the observed value $P_\zeta (k_\ast) = 2.2 \times 10^{-9}$ \cite{Ade:2013uln}. It determines the Hubble scale, $H_\ast$, when the pivot scale exited the horizon, and thus the energy density at that time, $\rho_\ast \simeq 3 M_P^2 H_\ast^2$, as a function of $f$ and $N_\ast$. Its dependence on $f$ and $N_\ast$ is mild, and we obtain $\rho_\ast \simeq 10^{16}$ GeV, see Fig.~\ref{fig:frho}.
On the other hand, from the Friedman equation for spatially flat Universe in the slow-roll approximation,
\begin{eqnarray}
3 H^2 M_P^2 = \rho \simeq V(\phi)\, ,
\label{Fried}
\end{eqnarray}
we obtain
\begin{eqnarray}
P_\zeta (k_\ast)
&\simeq& 
\frac{V(\phi_\ast)}{24\pi^2 M_P^4 \epsilon_{V\ast}}
\nn\\
&=&
\frac{c_0}{2\pi^2}\frac{1}{(2\pi L_5)^4}
\left[
1 - \cos \left( \frac{\phi_\ast}{f} \right)
\right]
\frac{1}{24\pi^2 M_P^4 \epsilon_{V\ast}}
\nn\\
&=&
\frac{c_0}{2\pi^2}
\frac{1}{(2\pi \ell_5)^6}
\left[
1 - \cos \left( \frac{\phi(f,N_\ast)}{f} \right)
\right]
\frac{1}{24\pi^2 \epsilon_V(\phi(f,N_\ast))} \, .
\label{Pzeta}
\end{eqnarray}
In the last line we have made it explicit that $\phi$ and $\epsilon_V$ are functions of $f$ and $N$.
Thus given $\ell_5$, $f$ and $N_\ast$, $c_0$ is determined from the observed value $P_\zeta (k_\ast) = 2.2 \times 10^{-9}$ by (\ref{Pzeta}).
The behaviour of $c_0$ as a function of $\ell_5$ is plotted in Fig.~\ref{fig:ellc0}. We observe that $c_0$ grows as $\ell_5^6$. Also, 
$c_0$ grows rapidly with $f$, as seen in Fig.~\ref{fig:fc0}.
Since each field charged under $U(1)_{B-L}$ with mass $\lesssim L_5^{-1}$ makes an $\Ord(1)$ contribution to $c_0$, if it is much larger than unity it may not be natural.\footnote{One can make large $c_0$ natural by introducing a large number in the model, e.g. a mupltiplet with a large multiplicity.}
Therefore we regard smaller values of $\ell_5$ and $f$, viz. $\ell_5\simeq 5$ and $f \simeq 5$, as preferred in our scenario here. With the independent parameters fixed, we then have $g_4 \simeq 10^{-3}$ and $L_5^{-1} \simeq 9 \times 10^{16}$~GeV from~(\ref{gbound}) and~(\ref{M5fg}). The value of the $U(1)_{B-L}$ gauge coupling is an important input to our $U(1)_{B-L}$ extension of the SM, which we discuss next.

\section{$U(1)_{B-L}$ extension of the Standard Model}\label{sec:BLSM}
There are several possibilities for the  $U(1)_{B-L}$ extension of the SM, particularly with regards to the charge assignment of the scalar field that would break the $U(1)_{B-L}$ symmetry. Table~\ref{charge} lists the particle content and the charge assignments of the particular $U(1)_{B-L}$ extension of the SM we consider here.
\begin{table}[ht]
\begin{center}
\begin{tabular}{|c|c|c|c|c|}
 \hline
 & $SU(3)_c$ & $SU(2)_L$ & $U(1)_Y$ & $U(1)_{B-L}$ \\
 \hline
 $q_L^i$ & ${\bf 3}$ & ${\bf 2}$ & $+1/6$ & $+1/3$ \\
 $u_R^i$ & ${\bf 3}$ & ${\bf 1}$ & $+2/3$ & $+1/3$ \\
 $d_R^i$ & ${\bf 3}$ & ${\bf 1}$ & $-1/3$ & $+1/3$ \\
 \hline
 $\ell_L^i$ & ${\bf 1}$ & ${\bf 2}$ & $-1/2$ & $-1$ \\
 $\nu_R^i$ & ${\bf 1}$ & ${\bf 1}$ & $0$ & $-1$ \\
 $e_R^i$ & ${\bf 1}$ & ${\bf 1}$ & $-1$ & $-1$ \\
 \hline
 $H$ & ${\bf 1}$ & ${\bf 2}$ & $-1/2$ & $0$ \\
 $\Sigma$ & ${\bf 1}$ & ${\bf 1}$ & $0$ & $+2$ \\
 \hline
\end{tabular}
\caption{Particle contents and charge assignment. The index $i = 1,2,3$ labels the generation.}
\label{charge}
\end{center}
\end{table}
In our set-up, we envisage all the SM particles and the right-handed neutrinos living on a four-dimensional brane, while the $U(1)_{B-L}$ gauge fields, $A_M$, and a complex scalar, $\Sigma$, responsible for the eventual $U(1)_{B-L}$breaking living in the five-dimensional bulk. In string theory this set-up may be realized, for example, when the SM fields and the right-handed neutrinos live on a (3+1)-dimensional D-brane localised in the extra dimension, while the bulk fields arise from higher dimensional D-branes.

The potential for the scalar sector renormalizable in four dimensions is given by
\begin{equation}
V(H,\Sigma) =
\mu_H^2 H^\dagger H + \mu_\Sigma^2 \Sigma_0^\ast \Sigma_0
+ \frac{\lambda_1}{2} (H^\dagger H)^2
+ \frac{\lambda_2}{2} (\Sigma_0^\ast \Sigma_0)^2
+ \lambda_3 H^\dagger H \Sigma_0^\ast \Sigma_0 \,.
\label{Hsect}
\end{equation}
Here, $\Sigma_0$ is the zero-mode of $\Sigma$ in the fifth direction.
After spontaneous symmetry breaking, the scalar fields acquire vacuum expectation values (VEVs), and we can write
\begin{equation}
H = \frac{1}{\sqrt{2}} 
\begin{pmatrix}
0 \\
v_H + h
\end{pmatrix} , \qquad
\Sigma_0 = \frac{v_\Sigma + s}{\sqrt{2}} \,,
\end{equation}
where $h$ and $s$ are excitations about the minimum, which is given by
\begin{equation}
\frac{v_H^2}{2} = \frac{-\mu_H^2\lambda_2 + \mu_\Sigma^2\lambda_3}{\lambda_1\lambda_2 - \lambda_3^2} \,, \qquad 
\frac{v_\Sigma^2}{2} = \frac{-\mu_\Sigma^2\lambda_1 + \mu_H^2\lambda_3}{\lambda_1\lambda_2 - \lambda_3^2} \,.
\end{equation}
Note that the $W$ boson mass fixes $v_H = 246$~GeV. 
In terms of $h$ and $s$, the quadratic part of the potential is given by
\begin{equation}
V^{(2)} = \frac{1}{2}\,\eta^\intercal M_0^2\,\eta \,, \quad 
\eta = 
\begin{pmatrix}
h \\
s
\end{pmatrix} ,
\end{equation}
where 
\begin{equation}
M_0^2 = 
\begin{pmatrix}
\lambda_1 v_H^2 & \lambda_3 v_H v_\Sigma \\
\lambda_3 v_H v_\Sigma & \lambda_2 v_\Sigma^2
\end{pmatrix} ,
\end{equation}
is the tree-level mass-squared matrix for $h$ and $s$, and we have used the minimization condition for the potential.
Diagonalizing, the physical mass eigenstates are defined by
\begin{equation}
\begin{pmatrix}
h \\
s
\end{pmatrix}
=
\begin{pmatrix}
 \cos\alpha & \sin\alpha \\
-\sin\alpha & \cos\alpha
\end{pmatrix}
\begin{pmatrix}
h_1 \\
h_2
\end{pmatrix} ,
\end{equation}
with the mixing angle given by
\begin{equation}
\tan 2\alpha = \frac{2\lambda_3 v_h v_s}{\lambda_2 v_s^2 - \lambda_1 v_h^2 } \,.
\end{equation}
The masses of the physical states are then given by
\begin{equation}
m_{h_{1,2}}^2 = \frac{1}{2}\left\{\lambda_1 v_H^2 + \lambda_2 v_\Sigma^2
\mp\sqrt{(\lambda_1 v_H^2 - \lambda_2 v_\Sigma^2)^2 + 4\lambda_3^2 v_H^2 v_\Sigma^2}\right\} \,.
\end{equation}
For $|\lambda_3 | \ll 1$ and $|v_H/v_\Sigma| \ll 1$, we can expand the square root and obtain
\begin{equation}
m_{h_{1,2}}^2 = \lambda_{1,2}v_{H,\Sigma}^2\mp\frac{\lambda_3^2}{\lambda_2}v_H^2 + \lambda_3^2 v_H^2\,\mathcal{O}\!\left(\frac{v_H^2}{v_\Sigma^2}\right) \,.
\end{equation}

\begin{figure}[htbp]
\centering
\includegraphics[width=2.6in]{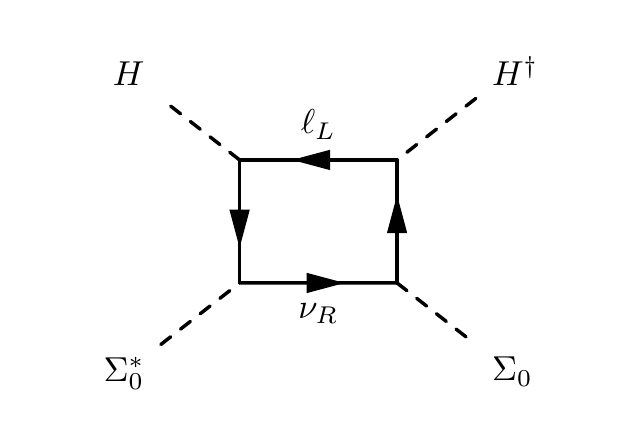} 
\caption{\label{fig:box}One-loop diagram 
which contributes to the mixing term 
$H^\dagger H \Sigma_0^\ast\Sigma_0$
through the right-handed neutrinos.}
\end{figure}

Assuming no coupling between the Higgs $H$ and the scalar $\Sigma_0$ at tree level, 
the mixing term $H^\dagger H \Sigma_0^\ast \Sigma_0$ is induced 
at one-loop level \cite{Iso:2009ss}
through interactions with neutrinos responsible for the seesaw mechanism 
\cite{Minkowski:1977sc,Yanagida:1979as,GellMann:1980vs,Glashow:1979nm}:
\begin{equation}
\mathcal{L}\supset -Y_D^{ij}\,\overline{\nu_R^i}\,H^\dagger\,l_L^j 
- \frac{1}{2}\,Y_N^{ij}\,\overline{\nu_R^{ic}}\,\nu_R^j\,\Sigma_0 + \mathrm{h.c.} \,.
\end{equation}
Fig.~\ref{fig:box} displays the particular one-loop graph.\footnote{%
Contributions from two-loop diagrams studied in \cite{Iso:2009ss}
are suppressed in our model due to the smallness of the
$U(1)_{B-L}$ gauge coupling.}
After $U(1)_{B-L}$ symmetry breaking, the mixing term contributes to the Higgs mass is estimated as
\begin{equation}\label{eq:dmH2}
\delta m_H^2\sim\frac{Y_D^2 Y_N^2}{(4\pi)^2}\frac{v_\Sigma^2}{2}\sim\frac{m_\nu M_N^3}{(4\pi)^2 v_H^2} \,,
\end{equation}
where we have used the seesaw formula $m_\nu \sim Y_D^2 v_H^2/M_N$ with 
$M_N = Y_N v_\Sigma/\sqrt{2}$ being the mass of the right-handed neutrino. 
Given the observation of the Higgs boson with mass $126$ GeV at the LHC~\cite{Aad:2013wqa,Aad:2013xqa,CMS:xwa},
we should have $\sqrt{|\delta m_H^2|} \lesssim 100$~GeV if naturalness is a criterion. Thus if we take $m_\nu \sim 0.1$~eV, we have $M_N \lesssim 10^7$~GeV from~\eqref{eq:dmH2} and hence $v_\Sigma \lesssim 10^7/Y_N$~GeV. This translates to an upper bound on the mixing coupling
\begin{equation}
|\lambda_3|\sim\frac{Y_D^2 Y_N^2}{(4\pi)^2}\sim\frac{m_\nu M_N}{(4\pi)^2 v_H^2}Y_N^2 \lesssim 10^{-10}\,Y_N^2 \,.
\end{equation}
Assuming $Y_N \simeq \Ord(1)$, 
the mass of the physical $U(1)_{B-L}$ gauge boson is estimated as
\begin{eqnarray}
m_{Z'} \sim g_4 v_\Sigma \lesssim \Ord(10^4) \,\mathrm{GeV} \,.
\label{gbmass}
\end{eqnarray}
From collider experiments, one has $m_{Z'} \geq g_4 \times (6\,\mathrm{TeV})$ for a $U(1)_{B-L}$ $Z'$ boson~\cite{Carena:2004xs}.
Since $g_4 \lesssim 10^{-3}$, there are no stringent bounds on $m_{Z'}$.

\section{The inflaton decay}\label{sec:idecay}
The coupling between the inflaton and the SM particles is crucial at the time of reheating.
Let us first consider the following $\mathbb{Z}_2$ transformation:
\begin{eqnarray}
x^5 \rightarrow - x^5, \quad
A_5 \rightarrow - A_5\, . 
\label{reflection}
\end{eqnarray}
We choose the origin of the $x^5$ coordinate to be where the brane is localised. We assume there are no other fields with $\mathbb{Z}_2$-odd charges under (\ref{reflection}) that are lighter than $A_5$.
Then if this $\mathbb{Z}_2$ transformation is an exact symmetry, the inflaton is absolutely stable. This will be a problem, however,
since then the Universe could not be heated to bring forth the standard hot big bang cosmology. We therefore introduce a five-dimensional Chern-Simons term, which breaks the $\mathbb{Z}_2$ symmetry:
\begin{eqnarray}
S_{CS}
=
\frac{k}{48 \pi^3} \int  \,  \cA \cF^2,
\label{5CS}
\end{eqnarray}
where $\cA = \cA_M dx^M$, $\cF = \frac{1}{2} \cF_{MN} dx^M dx^N$, and $k$ is some integer. 
Here, $\cA_M$ is the $U(1)_{B-L}$ gauge field with mass dimension one, which is related to the canonically normalized fields by
\begin{eqnarray}
A_M^{(5)} &=& \frac{1}{g_5} \cA_M\, , \\
A_{\mu} 
&=& \frac{1}{g_4} \cA_{\mu\, 0}\, ,
\quad
\phi = \frac{1}{g_4} \cA_{5\, 0}\, ,
\label{AcA}
\end{eqnarray}
where $A_M^{(5)}$ is the $U(1)_{B-L}$ gauge field canonically normalized in five dimensions, $A_\mu$ that in four dimensions, and $\cA_{M\, 0}$ the zero-mode of $\cA_M$ in five dimensions.

The four-dimensional interaction of the zero-modes following from (\ref{5CS}) is
\begin{eqnarray}
\frac{k}{48\pi^3}
\int d^4x\,
(2 \pi L )
\frac{3}{4}
\epsilon^{\mu\nu\rho\sigma}
\cA_{5\, 0} \cF_{\mu\nu\, 0} {\cF}_{\rho\sigma\, 0}
&=&
\frac{k}{16\pi^2}
\int d^4x\,
\frac{\phi}{2\pi f}  \cF_{\mu\nu\, 0} \tilde{\cF}_{\, 0}^{\mu\nu} 
\nn\\
&=&
g_4^2
\frac{k}{16\pi^2}
\int d^4x\,
\frac{\phi}{2\pi f}  F_{\mu\nu} \tilde{F}^{\mu\nu} .
\label{BLaxion}
\end{eqnarray}
Here the subscript $0$ denotes that they are (made from) zero-modes in the fifth direction.

The coupling (\ref{BLaxion})
gives the dominant contribution to the
decay width at the tree level:
\begin{eqnarray}
\Gamma_{\phi \rightarrow AA}
\simeq
\frac{g_{4}^4}{16 \pi}\left(\frac{k}{32\pi^3}\right)^2 \frac{m_\phi^3}{f^2}
\,,
\label{idecayr}
\end{eqnarray}
where $m_\phi$ is the mass of the inflaton. As we have seen, $c_0$ is determined by (\ref{Pzeta}) once $f$ and $g_4$ are given. This then determines $m_\phi$:
\begin{eqnarray}
\frac{m_\phi^2}{2} 
=
\frac{V_0}{4f^2}
=
\frac{1}{4f^2} \frac{c_0}{\pi^2} \frac{1}{(2\pi L_5)^4}
=
\frac{c_0 g_4^4 f^2}{4\pi^2} \, .
\label{eq:mphi}
\end{eqnarray} 
The $U(1)_{B-L}$ gauge bosons decay to SM particles via the minimal couplings.
As this proceeds much faster than the inflaton decay, the reheating temperature
is governed by the inflaton decay width (\ref{idecayr}). It is estimated as
\begin{eqnarray}
T_R &=& \left(\frac{90}{\pi^2 g_\star(T_R)}\right)^{1/4}\sqrt{\Gamma M_P} 
\simeq \left(\frac{90}{g_\star(T_R)}\right)^{1/4}\frac{g_4^2}{4\pi}\frac{|k|}{32\pi^3}\sqrt{\frac{m_\phi^3 M_P}{f^2}}
\nn\\
&\simeq&  
| k | \times 1 \, \, \mathrm{GeV} \,,
\label{TR}
\end{eqnarray}
where in the last line, 
we have used the preferred values $f \simeq 5 M_P$ and $\ell_5 \simeq 5$, which gives 
$m_\phi \simeq 10^{13}$ GeV. The factor $g_\star(T)$ is the effective relativistic degrees of freedom at temperature $T$. For $T_R \simeq 1 \sim 10$ GeV, $g_\star(T_R) \simeq 60 \sim 90$.
From (\ref{TR}), the reheating temperature is much smaller than the $U(1)_{B-L}$ breaking scale given by $v_\Sigma \simeq \Ord(10^7)$~GeV, when $k$ is $\Ord(1-10)$.
Comparing (\ref{TR}) with the standard estimate of the number of e-folds~\cite{Lyth:2009zz}:
\begin{eqnarray}
N_\ast \simeq
49
+ \frac{2}{3} 
\ln \left( \frac{\rho_\ast^{1/4}}{10^{16}\, \mbox{GeV}} \right)
+ \frac{1}{3} \ln \left(\frac{T_R}{1\, \mbox{GeV}}\right) \, ,
\label{Ne}
\end{eqnarray}
we observe that $N_\ast \simeq 50$ is natural in our model, as advertised earlier.

\section{Summary and Discussions}\label{sec:summary}
In this letter, we have studied the interrelation between cosmology and particle physics in $U(1)_{B-L}$ extra-natural inflation with a gauged $U(1)_{B-L}$ extension of the SM localised on a brane. The cosmological observation constrains 
the value of the $U(1)_{B-L}$ gauge coupling to $g_4 \lesssim 10^{-3}$, which in turn constrains the particle physics scenario at high energy assuming naturalness.
On the other hand, with SM particles localised on a brane, allowed interaction between the inflaton and the SM particles are restricted. Together with the value of $g_4$, the decay width of the inflaton and the reheating temperature are determined.

By tuning of a few parameters or with some slight extension, our model may also be able to explain other cosmological observations such as the Baryon number asymmetry of the Universe and the dark matter abundance. Indeed, the right-handed neutrinos could play a role in the former through the leptogenesis.  
They are also dark matter candidates. Another possible dark matter candidate, which may be included in our model, is a light scalar field odd under the reflection of the extra dimension~(\ref{reflection}). These merit further investigations.

Our main purpose in this letter is to present an example in which the relation between the BSM physics and the inflation physics are specified, and theoretical and observational constraints on one side constrains the other. We discussed one example here, but there can be several other possibilities, even within gauged 
$U(1)_{B-L}$ extensions of the SM. For instance, one may put some of the SM fields in the bulk. It will be interesting to explore those related scenarios.

\vspace*{8mm}

\begin{center}
{\bf Acknowledgments}
\end{center}
The authors would like to thank 
Chong-Sun Chu, Satoshi Iso, Hiroshi Isono, Yoji Koyama and Chia-Min Lin
for stimulating discussions.
KF is grateful to his former institutions, 
National Center for Theoretical Sciences and
the Department of Physics, National Tsing-Hua University,
where part of this work has been done.

\bibliography{BLref}
\bibliographystyle{utphys}

\end{document}